\documentclass[pra,twocolumn]{revtex4}%
\usepackage{amsfonts}
\usepackage{amsmath}
\usepackage{amssymb}
\usepackage{graphicx}%
\begin{document}
\title{Vectorial Doppler complex spectrum and its application to the rotational detection}
\author{Shuxian Quan}
\affiliation{School of Mathematics and Physics, China University of Geosciences, Wuhan
430074, China}
\author{Ling Chen}
\email{lingchen@cug.edu.cn}
\affiliation{School of Mathematics and Physics, China University of Geosciences, Wuhan
430074, China}
\author{Siyao Wu}
\affiliation{School of Mathematics and Physics, China University of Geosciences, Wuhan
430074, China}
\author{Baocheng Zhang}
\email{zhangbaocheng@cug.edu.cn}
\affiliation{School of Mathematics and Physics, China University of Geosciences, Wuhan
430074, China}

\begin{abstract}
Vectorial polarized fields of light has been applied to detect the rotational
velocity by the rotational Doppler effect, but the measurement was made for
the rotation of a single-particle system. When the rotational surface is
rough, the scattered vectorial Doppler signal spectrum is complex. In this
paper, we make the complex spectrum analyses using orbital angular momentum
modal expansion method. It is found that the highest peak in the Fourier form
of the complex spectrum is obtained at the frequency shift $2l\Omega$ related
to the topological charge ($l$) of the incident vortex light and the
rotational velocity ($\Omega$) of the rough surface. Based on the complex
spectrum analysis, we construct a method to measure the magnitude and
direction of the rotational velocity simultaneously for a general object,
which has the practical application in remote sensing and astronomy.

\end{abstract}

\maketitle

\section{Introduction}

The concept of optical vortices was put forward firstly in 1989 \cite{CGR1989}%
, and soon after that it was shown that the optical vortices formed by
Laguerre-Gaussian (LG) beams had an orbital angular momentum (OAM) appearing
in the phase as $exp(il\phi)$, where $l$ is the topological charge (TC)
\cite{LA1992} and determines the OAM value ($l\hbar$) of the photon, and
$\phi$ is the azimuthal angle around the rotating axis. Through about 30 years
of development, the optical vortex has been applied to many different fields
such as optical communication, quantum information and so on \cite{SGY2019}.

Theoretically, it was suggested in 1996 that the optical vortex could be
applied to measure the rotation rate of a spinning object \cite{GDN1996}. When
the vortex beam is scattered by the rotating object, the frequency shift in
the OAM of the light is observed to deduce the rotation rate. The frequency
shift, in essence, is stemmed from the Doppler effect \cite{YC1964}, and this
phenomenon is also called the rotational Doppler effect (RDE)
\cite{BAG1981,BB1997,CRP1998,MJP2006,BTL2006,BT2011}. In 2013, the RDE was
first used for the detection of a spinning object in the experiment
\cite{LM2013,LSP2013}, where it was confirmed that the RDE works even for the
case that the angular momentum vector of the rotating object is parallel to
the observation direction. Since then, the RDE is also applied to detect the
rotation for the different macro and micro targets (see the review
\cite{CWL2022} and references therein). When the spinning object has a rough
surface, the RDE can still be used to measure the object's rotation rate. But
the spectrum for the scattered light by the rough surface is complex, the
analyses for the spectrum is significant for the measurement and it is usually
made using a method of the expansion of the different OAM components
corresponding to different values of TC
\cite{ZFZ2016,ZFZ2017,ASS2020,DRL2021,STR2022}.

All those above are based on scalar vortex beams in which the spatial
polarization has a fixed distribution, so the magnitude of the rotating
velocity can be, but the direction of the rotation cannot be determined in
those cases. A recent attractive study \cite{FWW2021,LLW2021} using the
cylindrical vectorial polarization fields (CVPFs)
\cite{QZ2009,DMK2018,ABK2019,CH2019} presented the simultaneous measurement
for the magnitude and direction of the velocity of a rotating particle. The
so-called CVPFs are characterized by spatially and cyclically variant
polarized fields across the transverse plane of the beams and polarization
vortices could be formed. In this recent study, the beam with the CVPFs was
illuminated on the rotating particle, and the relative phase difference
between the two beams was observed to determine the velocity vector (magnitude
and direction). However, they present only the measurement for the rotation of
a single particle mimicked by a digital micromirror device (DMD) (see also
other related measurements in Ref. \cite{RHT2014,WFW2022}), which is
equivalent to the result from the rotation of uniformly isotropic and
homogeneous surface. Whether it can be applied to measure the rotation of a
rough surface is unclear directly since the inhomogeneous property of the
rough surface could lead to complex Doppler speckle signals after the beams
were illuminated on the rotating rough surface.

In this paper, we study the measurement of the velocity vector in a transverse
plane normal to the light axis for a rotating tinfoil whose surface is rough
using the CVPFs. We make the theoretical analyses for the complex spectrum
scattered by the rotating rough surface, which has not been studied for the
CVPFs before. We also measure the Doppler frequency shift using a paraxial
two-path polarization detection method, in which the noises stemming from the
low frequency terms in the complex spectrum can be suppressed. Our results
show that the simultaneous measurement for the magnitude and direction of the
velocity in a transverse plane normal to the light axis is still feasible. The
extension of the vectorial Doppler metrology method to the general rotating
objects is significant in the actual detection and has the potential to apply
in remote sensing and astronomy.

\section{Scattered light}

We start with the consideration of a general CVPF, whose light field is
expressed using the Jones matrix as%
\begin{equation}
\overrightarrow{E}=\overrightarrow{E}_{0}\left(  r,z\right)  e^{-2\pi
ift}\left(
\begin{array}
[c]{c}%
\cos\left(  l\phi+\alpha\right)  \\
-\sin\left(  l\phi+\alpha\right)
\end{array}
\right)  ,
\end{equation}
which can be generated by the superposition of one right-hand polarized vortex
beam with TC $l$ and one left-hand polarized vortex beam with TC ($-l$). The
two beams have the same field amplitude $\overrightarrow{E}_{0}\left(
r,z\right)  $ at the overlapped position ($r,z$) and have the same frequency
$f$. For the CVPFs, $l$ is also understood as the order number of the
polarization which represents the rotating cycles of the polarization vector,
$\phi$ is the azimuthal angle, and $\alpha$ is the initial phase. It is easy
to see that the polarization of the CVPFs is dependent on spatial positions,
as presented in Fig. 1. Compare it with the field of scalar vortex light which
is expressed as $E_{l}=E_{0}(r,z)e^{il\theta}e^{-2\pi ift}$. In the transverse
direction perpendicular to the propagation direction of the vortex beam, the
field $E_{l}$ or its polarization is evenly distributed, as presented in Fig. 1.

\begin{figure}[ptb]
\centering
\includegraphics[width=1\columnwidth]{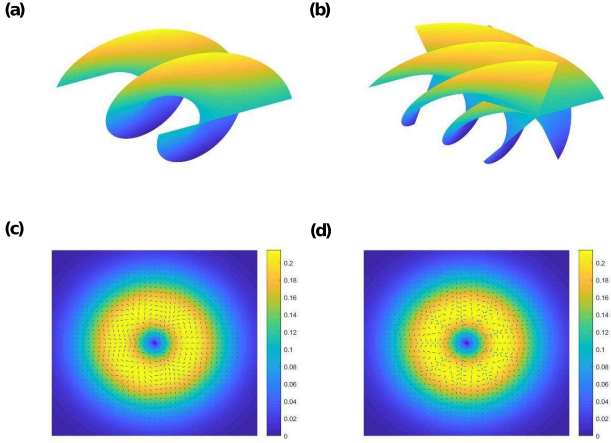}\caption{Schematic diagram of
optical vortices. The phase wavefront of scalar vortex beams are presented in
(a) for the TC $l=2$ and in (b) for the TC $l=5$. The polarization
distribution of vectorial vortex beams in the transverse plane are presented
in (c) for TC $l=2$ and in (d) for the TC $l=5$. }%
\label{Fig1}%
\end{figure}

When the vectorial beam is illuminated on a rough surface, its phase will be
modulated according to the modulation function \cite{PLS2014} as $M\left(
r,\phi,t\right)  =e^{i\varphi\left(  r,\phi\right)  }=\sum_{k}A_{k}\left(
r\right)  e^{ik\phi}$ where $A_{k}\left(  r\right)  $ is a complex amplitude
of $k$-order modulation factor and satisfies the normalization $\sum
_{k}\left\vert A_{k}\left(  r\right)  \right\vert ^{2}=1$. If the rough
surface is rotated with the angular frequency $\Omega$, the modulation
function becomes $M\left(  r,\Omega,t\right)  =$ $e^{i\varphi\left(
r,\theta-\Omega t\right)  }=\sum_{k}A_{k}\left(  r\right)  e^{ik\phi
}e^{-ik\Omega t}$.

After the vortex beam is scattered by the rough surface of the rotating body,
the light field becomes%
\begin{equation}
\overrightarrow{E}_{s}=M\left(  r,\Omega,t\right)  \overrightarrow{E},
\end{equation}
where $\overrightarrow{E}_{0}\left(  r,z\right)  e^{-2\pi ift}$ is ignored for
simplicity because only the change along the angular direction of the
scattered light field is required in the paper. It is calculated further as
\begin{equation}
\overrightarrow{E}_{s}=\overrightarrow{E}\left(  r\right)  \left(
\begin{array}
[c]{c}%
G\cos\left(  \alpha+l\Omega t\right)  +iF\sin\left(  \alpha+l\Omega t\right)
\\
-G\sin\left(  \alpha+l\Omega t\right)  +iF\cos\left(  \alpha+l\Omega t\right)
\end{array}
\right)  ,
\end{equation}
which can be divided into the superposition of two beams with different
polarizations,%
\begin{equation}
\overrightarrow{E}_{s}=G\left(
\begin{array}
[c]{c}%
\cos\left(  \alpha+l\Omega t\right)  \\
-\sin\left(  \alpha+l\Omega t\right)
\end{array}
\right)  +iF\left(
\begin{array}
[c]{c}%
\sin\left(  \alpha+l\Omega t\right)  \\
\cos\left(  \alpha+l\Omega t\right)
\end{array}
\right)  ,\label{tdp}%
\end{equation}
where $\Omega$ is the angular frequency of the rotating body. The modulation
functions $G=%
{\displaystyle\sum\limits_{m=-N}^{N}}
\left(  A_{m-l}+A_{m+l}\right)  e^{-im\Omega t}e^{im\phi}$, $F=%
{\displaystyle\sum\limits_{m=-N}^{N}}
\left(  A_{m-l}-A_{m+l}\right)  e^{-im\Omega t}e^{im\phi}$ where $m$
represents the OAM mode of the scattered light, and $m=k+l$. $G$ can be
regarded as the superposition of different harmonic oscillations with the
frequency $m\Omega$ and amplitude $A_{m-l}+A_{m+l}$, and $F$ can be regarded
as the superposition of different harmonic oscillations with the frequency
$m\Omega$ and amplitude $A_{m-l}-A_{m+l}$ but along the direction
perpendicular to that for $G$. According to Eq. (\ref{tdp}), the scattered
field can be obtained by a rotation at the angle $l\Omega t$ (in our paper the
counterclockwise direction is defined as positive) for the CVPFs with the form
$\left(  G,F\right)  ^{T}$ whose polarization distribution is not fixed and
changes with time. In particular, if the angular frequency $\Omega$ remains
unchanged but the direction of rotation is reversed (i.e. $\Omega$ becomes
$-\Omega$ in Eq. (\ref{tdp})), the scattered field will be changed, i.e.
$\overrightarrow{E}_{s}\left(  -\Omega,t\right)  \neq$ $\overrightarrow{E}%
_{s}\left(  \Omega,t\right)  $, which shows that the polarization distribution
in the scattered CVPFs depends on the direction of the rotation. This provides
a method to detect the direction of a rotation.

When the scattered light goes through a polarizer with the angular $\theta$
relative to the horizontal direction, the light field becomes%
\begin{align}
\overrightarrow{E}_{sf}  &  =\left[
\begin{array}
[c]{cc}%
\cos\theta & \sin\theta\\
0 & 0
\end{array}
\right]  \overrightarrow{E}_{s}\nonumber\\
&  =G\left[
\begin{array}
[c]{c}%
\cos\left(  l\Omega t+\theta\right) \\
0
\end{array}
\right]  +iF\left[
\begin{array}
[c]{c}%
\sin\left(  l\Omega t+\theta\right) \\
0
\end{array}
\right]  , \label{ppd}%
\end{align}
where the initial phase $\alpha=0$ is taken for simplicity. The light
intensity is obtained as%
\begin{align}
I_{j}  &  =\left\vert G\right\vert ^{2}\cos^{2}\left(  l\Omega t+\theta
\right)  +\left\vert F\right\vert ^{2}\sin^{2}\left(  l\Omega t+\theta\right)
\nonumber\\
&  +\frac{1}{2}i(G^{\ast}F-GF^{\ast})\sin2(l\Omega t+\theta),\nonumber\\
&  =\frac{1}{2}\left(  \left\vert G\right\vert ^{2}-\left\vert F\right\vert
^{2}\right)  \cos2(l\Omega t+\theta)\nonumber\\
&  +\frac{\left\vert G\right\vert ^{2}+\left\vert F\right\vert ^{2}}{2}%
+\frac{1}{2}i(G^{\ast}F-GF^{\ast})\sin2(l\Omega t+\theta), \label{li}%
\end{align}
where $\ast$ represents the complex conjugation, and the relations $\cos
^{2}\left(  l\Omega t+\theta\right)  =\frac{1+\cos2(l\Omega t+\theta)}{2}$ and
$\sin^{2}\left(  l\Omega t+\theta\right)  =\frac{1-\cos2(l\Omega t+\theta)}%
{2}$ are used. Although the last term has the imaginary number $i$ in the
result, but the term $i(G^{\ast}F-GF^{\ast})$ is real, as calculated below.

To simplify the expression of the intensity, we have to calculate the terms
related to modulation functions. At first, we calculate the term $\left\vert
G\right\vert ^{2}$ as%
\begin{align}
\left\vert G\right\vert ^{2} &  =GG^{\ast}\nonumber\\
&  =\sum_{m,m^{\prime}}\left(  A_{m-l}+A_{m+l}\right)  \left(  A_{m^{\prime
}-l}^{\ast}+A_{m^{\prime}+l}^{\ast}\right)  e^{i\left(  m-m^{\prime}\right)
\left(  \phi+\Omega t\right)  }.\label{m1}%
\end{align}
It can be seen that the right term is real by taking $m-m^{\prime}=n$ where
$n$ is an integer in the range from $0$ to $2m$. When $m-m^{\prime
}=n,\left\vert G\right\vert ^{2}=\sum_{m}\left(  A_{m-l}+A_{m+l}\right)
\left(  A_{m+n-l}^{\ast}+A_{m+n+l}^{\ast}\right)  e^{in\left(  \phi+\Omega
t\right)  }$; when $m-m^{\prime}=-n,$ $\left\vert G\right\vert ^{2}=\sum
_{m}\left(  A_{m+n-l}+A_{m+n+l}\right)  \left(  A_{m-l}^{\ast}+A_{m+l}^{\ast
}\right)  e^{-in\left(  \phi+\Omega t\right)  }$. These two terms are complex
conjugated, so their sum is real. Since the $n$ term and the ($-n$) term
appear with pairs and they are complex conjugated, the total sum are real.

Similarly, we can calculate the term $\left\vert F\right\vert ^{2}$ as%
\begin{align}
\left\vert F\right\vert ^{2} &  =FF^{\ast}\nonumber\\
&  =\sum_{m,m^{\prime}}\left(  A_{m-l}-A_{m+l}\right)  \left(  A_{m^{\prime
}-l}^{\ast}-A_{m^{\prime}+l}^{\ast}\right)  e^{i\left(  m-m^{\prime}\right)
\left(  \phi+\Omega t\right)  }.\label{m2}%
\end{align}
It is also real. Then, we calculate
\begin{align}
&  \frac{1}{2}\left(  \left\vert G\right\vert ^{2}-\left\vert F\right\vert
^{2}\right)  \nonumber\\
&  =\sum_{m,m^{\prime}}(A_{m-l}A_{m^{\prime}+l}^{\ast}+A_{m+l}A_{m^{\prime}%
-l}^{\ast})e^{i\left(  m-m^{\prime}\right)  \left(  \phi+\Omega t\right)
},\nonumber\\
&  =\sum_{m,n}[(A_{m-l}A_{m+n+l}^{\ast}+A_{m+l}A_{m+n-l}^{\ast})e^{in\left(
\phi+\Omega t\right)  }+c.c.],\label{mm}%
\end{align}
where $c.c.$ represents the complex conjugation of the former terms. Now we
write these expressions in the actual form. At first, we analyze such sum:
$(a+bi)e^{i\phi}+(a-bi)e^{-i\phi}=2a\cos\phi-2b\sin\phi=2\sqrt{a^{2}+b^{2}%
}(\frac{a}{\sqrt{a^{2}+b^{2}}}\cos\phi-\frac{b}{\sqrt{a^{2}+b^{2}}}\sin
\phi)=2\sqrt{a^{2}+b^{2}}(\cos\phi^{\prime}\cos\phi-\sin\phi^{\prime}\sin
\phi)=2\sqrt{a^{2}+b^{2}}\cos(\phi^{\prime}+\phi)$ where $\sqrt{a^{2}+b^{2}}$
and $\phi^{\prime}$ are the modulus and argument of the complex number
$(a+bi)$. Since the summation terms in Eqs. (\ref{m1}) and (\ref{m2}) are
complex conjugated pairwise, thus according to the analysis above we obtain%
\begin{align}
&  \frac{1}{2}\left(  \left\vert G\right\vert ^{2}-\left\vert F\right\vert
^{2}\right)  \nonumber\\
&  =2\sum_{m,n}(\left\vert A_{m-l}\right\vert \left\vert A_{m+n+l}^{\ast
}\right\vert \cos(n(\phi+\Omega t)+\varphi_{n}^{-l})\nonumber\\
&  +\left\vert A_{m+l}\right\vert \left\vert A_{m+n-l}^{\ast}\right\vert
\cos(n(\phi+\Omega t)+\varphi_{n}^{l})),\label{p1}%
\end{align}
where $\varphi_{n}^{-l}$ and $\varphi_{n}^{l}$ are the arguments of the
complex numbers $A_{m-l}A_{m+n+l}^{\ast}$ and $A_{m+l}A_{m+n-l}^{\ast}$. Using
the same analyses, we obtain
\begin{align}
&  \frac{1}{2}\left(  \left\vert G\right\vert ^{2}+\left\vert F\right\vert
^{2}\right)  \nonumber\\
&  =2\sum_{m,n}(\left\vert A_{m-l}\right\vert \left\vert A_{m+n-l}^{\ast
}\right\vert \cos(n(\phi+\Omega t)+\psi_{n}^{-l})\nonumber\\
&  +\left\vert A_{m+l}\right\vert \left\vert A_{m+n+l}^{\ast}\right\vert
\cos(n(\phi+\Omega t)+\psi_{n}^{l})),\label{p2}%
\end{align}
where $\psi_{n}^{-l}$ and $\psi_{n}^{l}$ are the arguments of the complex
numbers $A_{k-l}A_{k+p+l}^{\ast}$ and $A_{k+l}A_{k+p-l}^{\ast}$, and
\begin{align}
&  \frac{1}{2}i(G^{\ast}F-GF^{\ast})\nonumber\\
&  =-2\sum_{m,n}(\left\vert A_{m-l}\right\vert \left\vert A_{m+n+l}^{\ast
}\right\vert \sin(n(\phi+\Omega t)+\varphi_{n}^{-l})\nonumber\\
&  +\left\vert A_{m+l}\right\vert \left\vert A_{m+n-l}^{\ast}\right\vert
\sin(n(\phi+\Omega t)+\varphi_{n}^{l})).\label{p3}%
\end{align}
Then substitute Eqs. (\ref{p1})-(\ref{p3}) into the expression (\ref{li}) of
light intensity, and we obtain%
\begin{align}
I &  =\left\vert \overrightarrow{E}_{sf}\right\vert ^{2}=2\sum_{n=0}^{N}%
\sum_{m=-Z}^{Z-n}[\left\vert A_{m+l}\right\vert \left\vert A_{m+n-l}^{\ast
}\right\vert \nonumber\\
&  \times\cos(\left(  2l+n\right)  \Omega t+n\phi+2\theta+\varphi_{n}%
^{l})\nonumber\\
&  +\left\vert A_{m-l}\right\vert \left\vert A_{m+n+l}^{\ast}\right\vert
\cos(\left(  2l-n\right)  \Omega t-n\phi+2\theta-\varphi_{n}^{-l})\nonumber\\
&  +\left\vert A_{m+l}\right\vert \left\vert A_{m+n+l}^{\ast}\right\vert
\cos(n\Omega t+n\phi+\psi_{n}^{l})\nonumber\\
&  +\left\vert A_{m-l}\right\vert \left\vert A_{m+n-l}^{\ast}\right\vert
\cos(n\Omega t+n\phi+\psi_{n}^{-l})].\label{lif}%
\end{align}
When $n=0$, the last two term becomes $\sum_{m}\left\vert A_{m+l}\right\vert
^{2}+\sum_{m}\left\vert A_{m-l}\right\vert ^{2}$ which are independent on the
rotational angular frequency. $n$, $N$, and $Z$ are positive integers. $Z$ is
the detectable maximal TC, and the detectable TCs are $0$, $\pm1$, $\pm2$,
$\cdots$, $\pm Z$ for the symmetrical measurement. $N=2Z$ represents the modes
of beat frequency, which means the maximal difference between two arbitrary
modes from $-Z$ to $Z$. For a given $n$-order harmonic component, there exist
($N-n+1$) terms contributed to the intensity. An illuminated interpretation is
presented in Fig. 2.

The intensity $I$ consists of four terms, but only the former two terms are
related to the angle of polarizer. This provides the possibility to detect the
rotation velocity by changing the angle of the polarizer. We called them beat
frequency terms, which include different beat frequencies with the central
frequency $2l\Omega/2\pi$, i.e. $2l\Omega/2\pi$, $\left(  2l\pm1\right)
\Omega/2\pi$, $\left(  2l\pm2\right)  \Omega/2\pi$, $\left(  2l\pm3\right)
\Omega/2\pi$, $\cdots$. The other two terms are called as low frequency terms.
Theoretically, the scattered light should contain all possible modes, but in
the practical experiment, only a small quantity of modes (i.e. $N$ is small)
can be measured. In particular, the amplitude of the mode with frequency
$2l\Omega/2\pi$ is maximal, as seen in Fig. 2.

\begin{figure}[ptb]
\centering
\includegraphics[width=1\columnwidth]{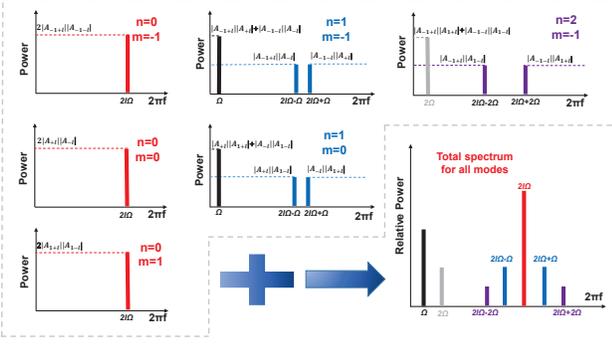}\caption{Schematic diagram of
complex spectrum with $Z=1,N=2$ in Eq. (\ref{lif}). The figures in the same
row are presented with the same scattering mode $m$ but different harmonic
order $n$, and the figures in the same column are presented with the same
harmonic order $n$ but different scattering mode $m$. Here we assume that the
modulated amplitudes of scattered light by the rough surface are the same. The
figure in the lower right corner represents the complex spectrum by summing
all these components, which gives the maximal amplitude at the frequency
$2l\Omega$. }%
\label{Fig2}%
\end{figure}

From the calculation above, it is seen that the distribution for the amplitude
and direction of the polarization depends on the angle position $\phi$ in the
transverse plane perpendicular to the propagation direction of the light.
Meanwhile, in the process of light propagation, the distribution is rotated
with a frequency $l\Omega$, and the rotation chirality for the polarization
distribution will be changed if the direction of the rotating rough surface is
changed. These provide the basis for detecting the amplitude and direction of
the rotating objects.

A particular operation has to be stressed here. As calculated in Eq.
(\ref{ppd}), a polarizer is added before the detection for the scattered
light. If the polarizer is not added, the detected light intensity is
$I=E^{\ast}E=G^{2}+F^{2}$ which includes only some low frequency terms
unrelated to the TC of the incident light, similar to the last two terms in
Eq. (\ref{lif}). This cannot provide the method to detect the direction of the
rotation. When the polarizer is added, the beat frequency terms appear in the
final expression for the light intensity as given in the former two terms in
Eq. (\ref{lif}), which makes the detection of the direction possible. This is
different from the scalar light field. If the incident light is scalar, the
scattered light is also scalar. Thus, even if the polarizer is added before
the detection, only the amplitude of the scalar light is modulated (i.e.
$I\propto f\left(  \theta\right)  \cos\left(  2l\Omega t+\alpha\right)  $,
$\theta$ is the angle of the added polarizer and $f\left(  \theta\right)  $ is
a function related to the modulated amplitude) and so the direction cannot be
detected in this case.

Then, a question is how to find the correspondence between the peak of the
light intensity curve and the frequency since the scattered light field is so
complex. At first, we explain the relationship between the parameters $A$, $m$
and the rough degree of the rough surface. $m$ is determined by the height
from the reference plane to the surface of the material, and $A$ is determined
by the surface area of the material at the same height. In our experiment, the
fluctuation of the height in the surface of the tinfoil is at the micron
scale, which lead to the small change for $A$ and $m$ in the different
positions. So in the expansion of Eq. (\ref{lif}), the parameters $\left\vert
A_{m}\right\vert $ have the approximately same value. Thus, we can estimate
which frequency corresponds to the highest peak in the Fourier form of the
spectrum by a simple calculation using Eq. (\ref{lif}). In our experiment, we
use the paraxial symmetric measurement, in which the symmetric modes such as
$0$, $\pm1$, $\pm2$ and so on are collected into the detector. Generally, when
the maximal detectable mode is $m$, there are $2m$ terms for the low frequency
modes, and there are $4m+1$ terms for the beat frequency modes with the
central frequency $2l\Omega$. Note that the contribution for the mode of
frequency $2l\Omega$ is from $2m+1$ terms and every term contributes to the
amplitude by $2A^{2}$, and thus, the total contribution is $(4m+2)A^{2}$,
which leads to the highest peak in the spectrum curve. An illuminated
interpretation is presented in Fig. 2 with the maximal detectable mode is $1$.
In this figure, every row represents the beat frequencies between the same
mode with all possible modes, i.e. the first row presents the beat frequencies
between the mode with $m=-1$ and the modes with $m=-1,0,1$ which corresponds
to the harmonic orders $n=0,1,2$; the second row presents the beat frequencies
between the mode with $m=0$ and the modes with $m=0,1$ which corresponds to
the harmonic orders $n=0,1$; the third row presents the beat frequencies
between the mode with $m=1$ and the mode with $m=1$ which corresponds to the
harmonic order $n=0$. According to Eq. (\ref{lif}), the harmonic order $n=0$
corresponds to the peak at the frequency $2l\Omega$, and there are three terms
contributed to the intensity with every one $2A^{2}$. The intensity at other
frequency can be counted with such a method, as presented in the final plot
(Total spectrum for all modes) in Fig. 2.

\begin{figure}[ptb]
\centering
\includegraphics[width=1\columnwidth]{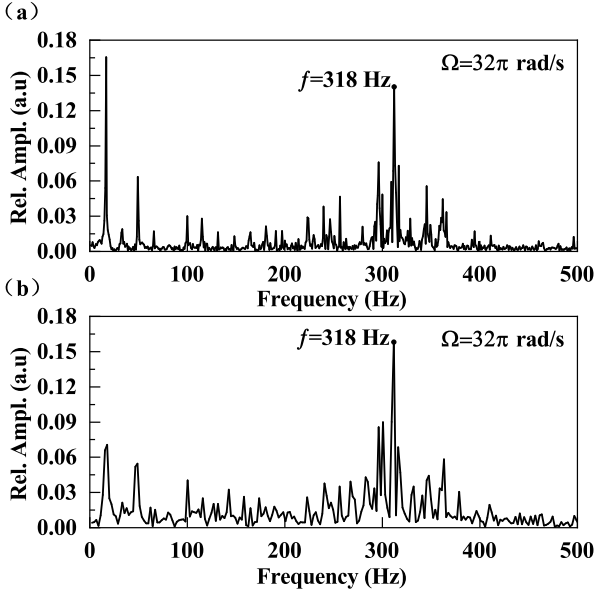}\caption{Fourier transformation
forms for measurement results of the complex spectrum with the angular
velocity of rotation, $\Omega=32\pi$ rad/s and the TC, $l=10$, set in advance
in our experiment. The down (up) plot indicates the results in which the low
frequency terms are (are not) subtracted. The subtraction is made using the
measurement results in the main path to subtract the results in the reference
path, and then the Fourier transformation is made.}%
\label{Fig3}%
\end{figure}

To confirm our theoretical analyses, Fig. 3a presents the Fourier transformed
form of the complex spectrum measured in the experiments, in which the signals
from the low frequency terms and the beat frequency terms in Eq. (\ref{lif})
can be found obviously. It is seen that the highest peak appears at the low
frequency $\Omega$, different from our theoretical estimation (the highest
peak should be at the frequency $2l\Omega/2\pi$, as analyzed in SM). This
could be caused by the noises from the mechanical vibration of the rotation
motor. In order to improve the measurement results, either the noises should
be subtracted using some specific methods, or the low frequency terms are
subtracted in the final measured results. Whichever one is realized, the
direct measurement for the light intensity is not feasible and some other
operations are required. In the next section, we will discuss the measurement
method and present the measurement process.

\section{Experimental setup}

\begin{figure}[ptb]
\centering
\includegraphics[width=1\columnwidth]{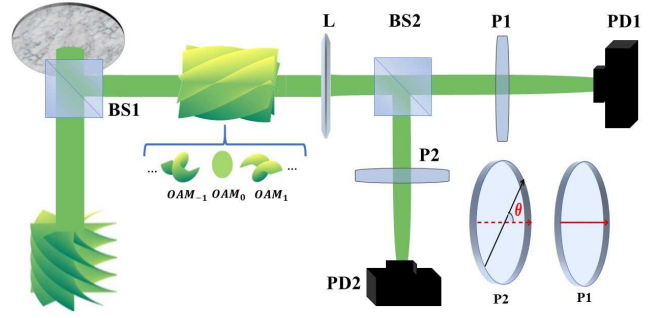}\caption{Measurement method.
The incident vectorial vortex light is scattered by the rotating rough
surface. The scattered light is divided by the beam splitter (BS) into two
paths. The path that the light goes through the polarizer (P1, its angle is
set at the zero degree for the whole experiment) is the reference path. The
path that the light goes through P2 (its angle will be changed according to
the requirement of the measurement) is called as the main path. }%
\label{Fig4}%
\end{figure}

\begin{figure}[ptb]
\centering
\includegraphics[width=1\columnwidth]{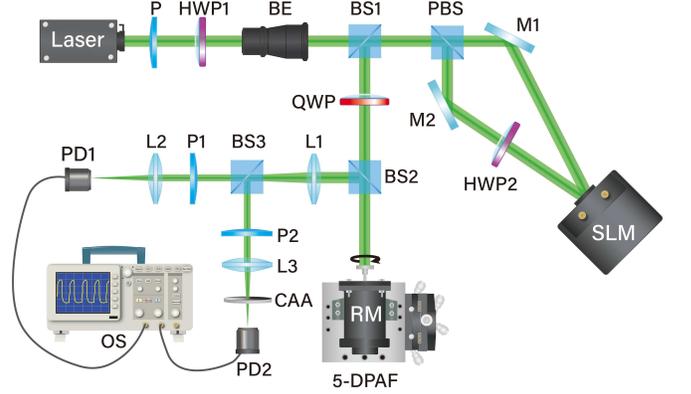}\caption{Experimental setup. P,
P1, P2: Polarizer; BE: Beam Expander; HWP1, HWP2: Half-Wave Plate; BS1, BS2,
BS3: Beam Splitter; PBS: Polarization Beam Splitter; QWP: Quarter-Wave Plate;
M1, M2: Mirror; SLM: Spatial Light Modulator; L1, L2, L3: Lens with the focal
length 100 mm; PD1, PD2: Photodetector; OS: Oscilloscope; CAA: Continuously
Adjustable Attenuator; 5-DPAF: Five Dimensional Precision Adjusting Frame; RM:
Rotation Motor.}%
\label{Fig5}%
\end{figure}

As analyzed above for the complex spectrum of the scattered light, we can
realize this by changing the angle $\theta$ of the polarizer. However, it is
noted that the modes in the low frequency terms are independent of the angle
$\theta$, so it has to be subtracted in the measurement. For this, we use a
beam splitter (BS) to divide the scattered light into two paths with the same
amplitude in each path, as presented in Fig. 4 (the complete experimental
setup is presented in Fig. 5, which contains three parts: the generation of
CVPFs, the scattering of CVPFs by the rotating rough surface, and the
detection of the scattering light). The path including a polarizer (P1) is
regarded as the reference path and the angle of P1 is set at zero degree for
the whole experiments. The path including the polarizer (P2) is called as the
main path, and the measurement is realized by changing the angle of P2. For
the measurement to obtain the magnitude of the rotational velocity, we can use
the light intensity ($I_{2}\left(  \theta\right)  $) measured in the main path
to subtract the light intensity ($I_{1}\left(  \theta=0\right)  $) measured in
the reference path,
\begin{align}
I &  =I_{2}-I_{1}=4\sin\theta\sum_{n=0}^{N}\sum_{m=-Z}^{Z-n}[\left\vert
A_{m+l}\right\vert \left\vert A_{m+n-l}^{\ast}\right\vert \nonumber\\
&  \times\sin(\left(  2l+n\right)  \Omega t+n\phi+\theta+\varphi_{n}%
^{l})\nonumber\\
&  +\left\vert A_{m-l}\right\vert \left\vert A_{m+n+l}^{\ast}\right\vert
\sin(\left(  2l-n\right)  \Omega t-n\phi+\theta_{j}-\varphi_{n}^{-l}%
)],\label{lid}%
\end{align}
where Eq. (\ref{lif}) is used for the calculation. It is seen that the low
frequency terms have been reduced. Then, we find the frequency relative to the
maximal amplitude to deduce the value of the rotational velocity of the rough
surface using the relation $f=\frac{2l\Omega}{2\pi}$ for the known vortex beam
with the TC $l$. For the measurement to obtain the direction of rotation, we
can rotate the P2 in the main path along the same direction to increase the
angle $\theta$. The direction can be deduced by observing the shift of the
spectrum line, and the detailed discussion is given below.

From Eq. (\ref{lid}), it is deduced that the highest peak in the
frequency-domain spectrum curve corresponds to the beat frequency
$2l\Omega/2\pi$, which is presented in Fig. 3b, as expected in the paraxial
symmetrical measurement. It can be used to deduce the magnitude of rotational
velocity. According to our measurement, the value of rotational velocity is
31.8$\pi$ rad/s, which is nearly consistent with our setup (32$\pi$ rad/s) for
the rotation in advance. More similar cases are presented here, and all of
them show that the subtraction is necessary to extract properly the frequency
corresponding to the highest peak. Fig. 6 gives the results of the amplitude's
change with the frequency under the conditions that the rotation velocity is
the same, but the incident vectorial vortex beams are different with different
TCs. Fig. 7 gives similar results but under different conditions that the
incident vectorial vortex beams are the same and the rotation velocity is
different. The left column in Fig. 6 and Fig. 7 presents the results measured
in the main path, and the right column presents the difference of amplitude by
subtracting the results measured in the reference path from the results
measured in the main path. It is seen that the low-frequency terms are much
suppressed in the right column of Fig. 6 and Fig. 7. At the same time, it is
also seen that the highest peak is obtained at the frequency $2l\Omega$ in the
right column of Fig. 6 and Fig. 7. Actually, the results are applicable in any
other cases, but in our experiments, the rotation velocity cannot be set too
large due to the limitation of the devices.

\begin{figure}[ptb]
\centering
\includegraphics[width=1\columnwidth]{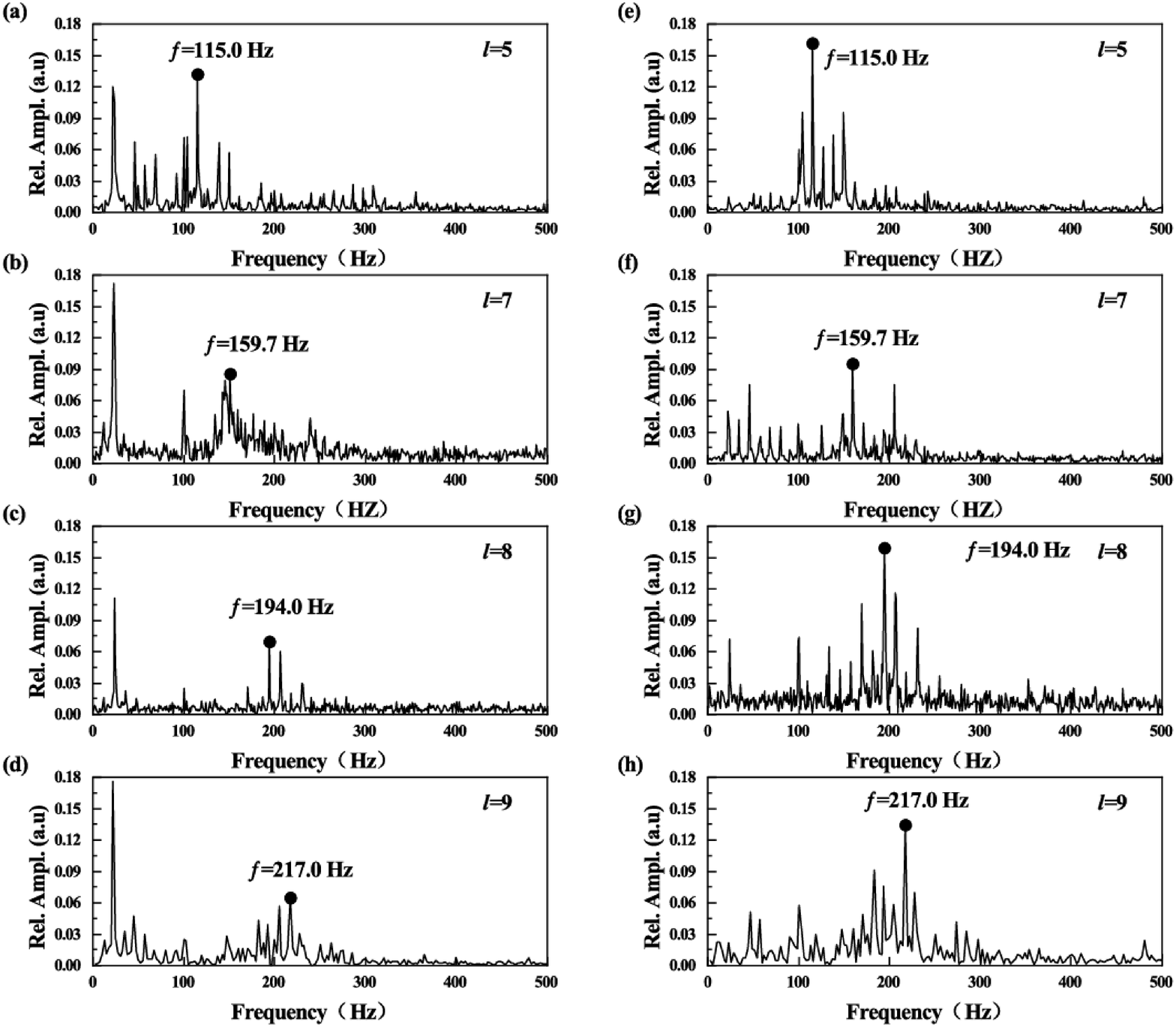}\caption{Experimental results
for the change of amplitude with the frequency under the same rotational
velocity of the rough surface with $\Omega=23\pi$ rad/s. The figures in (a-d)
present the measured results in the main path and the figures in (e-f) present
the difference between the measured results in the main path and those in the
reference path. }%
\label{Fig6}%
\end{figure}

\begin{figure}[ptb]
\centering
\includegraphics[width=1\columnwidth]{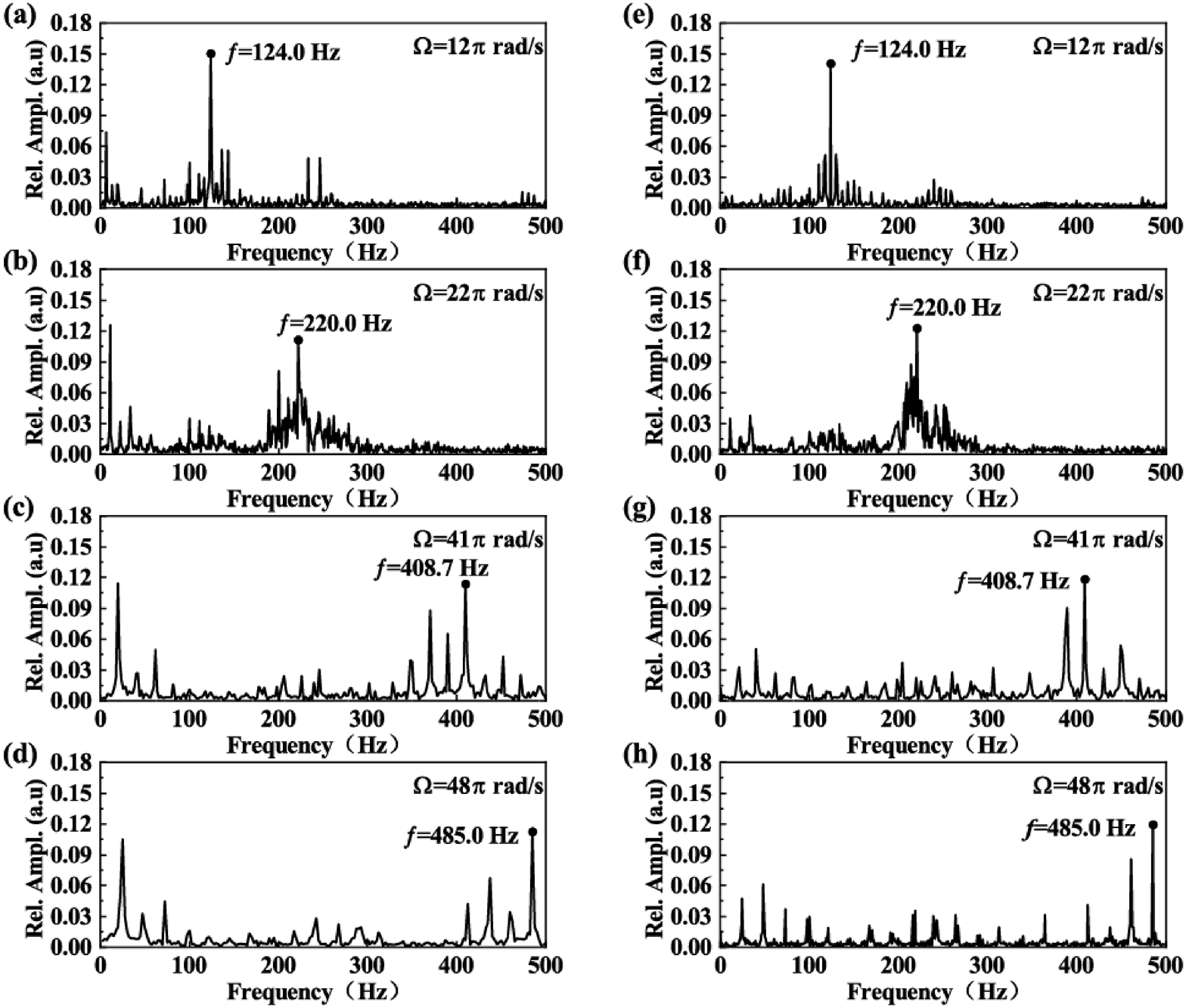}\caption{Experimental results
for the change of amplitude with the frequency under the same incident vortex
beams with TC $l=10$. The figures in (a-d) present the measured results in the
main path and the figures in (e-f) present the difference between the measured
results in the main path and those in the reference path. }%
\label{Fig7}%
\end{figure}

\section{Measurement results}

Fig. 8 presents the measurement results for the magnitude of the rotational
velocity. The data is obtained from the measurement of light intensity in the
main optical path after subtracting the values from the reference path. The
straight lines represent the theoretical results with the relation between the
measured frequency at the maximal amplitude and the angular velocity of
rotation, i.e. $f=2l\Omega/2\pi$. It is seen that the measurement matches the
theoretical results very well.
Actually, this can also be used to measure the TC of the optical vortex given
the rotational rate. Here we present the result of measuring the TC when the
angular velocity of rotation is known, as in Fig. 9. In our measurement, if
the rotation velocity is too small, the measurement will not be exact. This is
because the measured results of light intensity in the two optical paths
cannot be cancelled completely due to experimental errors. Thus, the low
frequency terms will disturb the experimental results for the low rotation velocity.

\begin{figure}[ptb]
\centering
\includegraphics[width=1\columnwidth]{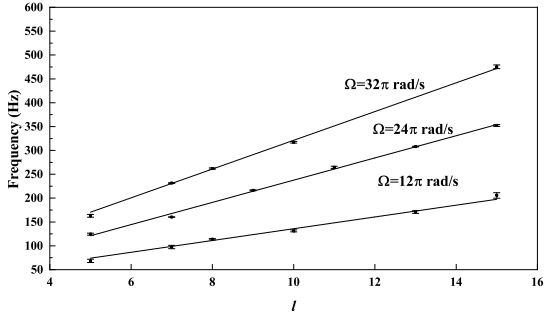}\caption{Measurement results
for the the angular velocity of rotation by the Doppler effect. The
measurement results are obtained from the frequency at which the amplitude is
maximal for every incident beam with TCs $l=5,7,8,9,10,11,13,15$. The error
bars stems mainly from the mechanical vibration. The angular velocity of
rotation is obtained by $\pi$ times the slope of the lines.}%
\label{Fig8}%
\end{figure}

\begin{figure}[ptb]
\centering
\includegraphics[width=1\columnwidth]{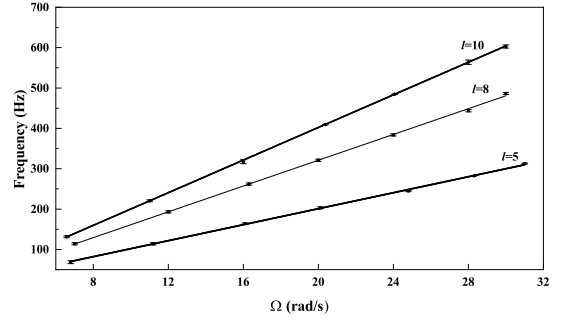}\caption{Experimental results
for the measurement of TC. The measurement results are obtained from the
frequency at which the amplitude is maximum for different rotational velocity.
The data points are obtained from the measurement results in the main path but
subtracting the results in the reference path. They matches well with the
theoretical results represented by the straight line. The TC can be deduced
from the slope of the lines }%
\label{Fig9}%
\end{figure}

Fig. 10 describes how to measure the direction of rotation. The method is to
rotate the polarizer (P2) in the main optical path. From the left plot of Fig.
10, it is seen that the maximal amplitude for the beat frequency of the
$2l\Omega/2\pi$ component will shift toward the leftside when the polarizer is
rotated with the same direction with the detected rotation. Otherwise, when
the polarizer is rotated with the inverse direction with the detected
rotation, the measured maximal amplitude for the beat frequency of the
$2l\Omega/2\pi$ component will shift toward the rightside. Thus, we can
estimate the direction of rotation quickly and exactly by observing the shift
of these curved lines. In particular, the results in Fig. 10 is obtained from
the difference between the simultaneous measurements in the two paths. This
can subtract the noises from the low frequency terms and from time difference
for different angles of polarizer.

For the measurement of the rotational direction of the tinfoil, the
subtraction of the results measured in the reference path is necessary. This,
on one hand, can reduce the influence of the low frequency terms in the
complex spectrum, and on the other hand, it can suppress the influence of the
phase's errors due to the time difference for the measurements at different
angles of polarizer, since the change of the angle requires to cost time,
which leads to a new phase shift due to the time difference between two
measurements at two different angles. Fig. 11 presents the results obtained
only from the measurement in the main path without making the subtraction of
the results measured in the reference path. It is seen that the shift could
make mistakes for some changes in the angle of the polarizer. In fact, from
Fig. 10, it is seen that only two lines for two different angles of the
polarizer can lead us to deduce the rotational direction without any mistakes,
but from Fig. 11, the errors occur, which might lead to the false estimation
for the rotational direction due to the improper choices for the angles of the
polarizer. However, the observation for the shift of the spectrum line from
enough measurements at many different angles of the polarizer can also lead to
the right estimation for the rotational direction.

\begin{figure}[ptb]
\centering
\includegraphics[width=1\columnwidth]{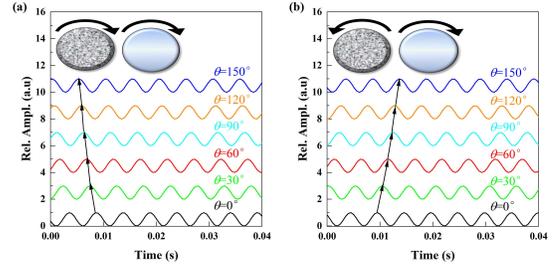}\caption{Measurement results
for the the direction of rotation. Signal modulation versus time are obtained
by subtracting the results in the reference path from that in the main path
and the different lines correspond to different angles of P2. The rough
surface and the P2 are presented with the same rotation direction in (a) and
the inverse direction in (b). The shift of spectra are marked with a line with
the arrow. }%
\label{Fig10}%
\end{figure}

\begin{figure}[ptb]
\centering
\includegraphics[width=1\columnwidth]{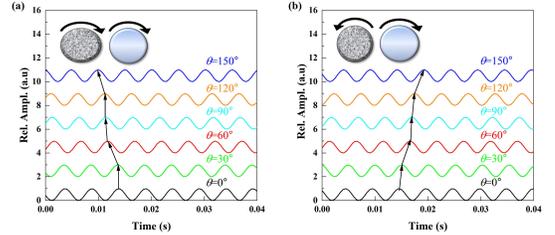}\caption{Measurement results
for the the direction of rotation. The change of amplitude with time are
obtained from the measurement results in the main path and the different lines
correspond to different angles of P2. The rough surface and the P2 are
presented with the same rotation direction in (a) and the inverse direction in
(b). The shift of spectra are marked with a line with the arrow. }%
\label{Fig11}%
\end{figure}

\section{Conclusions}

In this paper, we investigate the rotational Doppler effect based on the CVPFs
to detect the rotational velocity of a rough surface. We theoretically analyze
the Doppler OAM complex spectrum which consists of the scattered light by a
rough rotating surface, and obtain the expression for the intensity
distribution of scattered light using the OAM modal expansion method. The
highest peak in the Fourier form of the complex spectrum is found and it will
be shifted if the polarizer is rotated, which provides the method to measure
the magnitude and direction of the rotational velocity simultaneously. We also
implement an experimental measurement for this based on a paraxial two-path
method. One path is used as the reference to remove the noises related to the
low frequency term in the complex spectrum and the time delay at the
measurements for different angles of the polarizer in the main path. Different
from the earlier measurement using the single-mode fiber to filter the
specific mode, we use the paraxial symmetrical measurement to obtain the
complex spectrum within a small range of TC distribution. The highest peak and
its shift following the rotation of the polarizer are measured. This gives the
result about the rotational velocity, consistent with our theoretical
analyses. Thus, we expand the method of vectorial Doppler metrology to the
general situation, which goes forward a step to the practical application for
the detection of rotation.

\section{Acknowledgments}

We acknowledge the support from the fund of China University of Geosciences
(Wuhan). This work is also supported by the National Natural Science
Foundation of China (NSFC) under Grant No. 11654001.

\end{document}